\documentclass{sigchi}

\usepackage{soul}
\usepackage{times}
\usepackage{helvet}
\usepackage{courier}
\usepackage{xcolor}
\usepackage{graphicx}
\usepackage[hide]{chato-notes} 
\usepackage{caption}
\usepackage{subcaption}
\usepackage{float}
\usepackage{balance}
\usepackage{balance}  
\usepackage{url}      

\makeatletter
\def\url@leostyle{%
  \@ifundefined{selectfont}{\def\UrlFont{\sf}}{\def\UrlFont{\small\bf\ttfamily}}}
\makeatother
\def\pprw{8.5in}
\def\pprh{11in}

\toappear{\scriptsize Permission to make digital or hard copies of all or part of this work for personal or classroom use is granted without fee provided that copies are not made or distributed for profit or commercial advantage and that copies bear this notice and the full citation on the first page. Copyrights for components of this work owned by others than ACM must be honored. Abstracting with credit is permitted. To copy otherwise, or republish, to post on servers or to redistribute to lists, requires prior specific permission and/or a fee. Request permissions from permissions@acm.org. \\
{\emph{CHI 2015}}, April 18--23, 2015, Seoul, Republic of Korea. \\
Copyright \copyright~2015 ACM 978-1-4503-3145-6/15/04\ ...\$15.00. \\
http://dx.doi.org/10.1145/2702123.2702153}
\begin{document}

\title{You Tweet What You Eat:\\Studying Food Consumption Through Twitter\vspace{0.5cm}}
\author{ Sofiane Abbar, Yelena Mejova, Ingmar Weber\\
\affaddr{Qatar Computing Research Institute, PO 5825, Doha, Qatar}\\
\email{\{sabbar,ymejova,iweber\}@qf.org.qa}\\
\vspace{0.5cm}
}

\maketitle

\begin{abstract}
\begin{quote}
Food is an integral part of our lives, cultures, and well-being, and is of major interest to public health. 
The collection of daily nutritional data involves keeping detailed diaries or periodic surveys and is limited in scope and reach. 
Alternatively, social media is infamous for allowing its users to update the world on the minutiae of their daily lives,
including their eating habits. 
In this work we examine the potential of Twitter to provide insight into US-wide dietary choices by linking the tweeted dining 
experiences of $210$K users to their interests, demographics, and social networks. 
We validate our approach by relating the caloric values of the foods mentioned in the tweets to the state-wide obesity rates, achieving a Pearson 
correlation of $0.77$ across the 50 US states and the District of Columbia. 
We then build a model to predict county-wide obesity and diabetes statistics based on a combination of demographic variables and food names mentioned on Twitter. Our results show significant improvement over previous CHI research \cite{Culotta14}. We further link this data to societal and 
economic factors, such as education and income, illustrating that areas with higher education levels tweet about 
food that is significantly less caloric. 
Finally, we address the somewhat controversial issue of the social nature of obesity (Christakis \& Fowler \cite{christakis2007spread}) by 
inducing two social networks using mentions and reciprocal following relationships. 
\end{quote}
\end{abstract}



\section{Introduction}
\label{sec:introduction}

Food is a part of our daily lives that determines our well-being, health, and longevity. It is an important social activity and an expression of our culture and beliefs. The study of dietary habits is important for both cultural understanding and for monitoring public health. Heart disease, diabetes, osteoarthritis, and even cancer have all been linked to weight gain\footnote{\url{http://www.nhlbi.nih.gov/health/health-topics/topics/obe/risks.html}} and the US Center for Disease Control and Prevention (CDC) estimates some 35.7\% of adults in US are obese\footnote{\url{http://www.cdc.gov/obesity/data/facts.html}}, with medical care and other expenses associated with obesity costing up to \$$190$ billion a year \cite{cawley2012medical}. To best address this issue, public health awareness campaigns use data on dietary behavior across various segments of US population to tailor their messages to particular focus groups \cite{donohew1990public}. 
Having detailed and accurate data on the cultural and individual behaviors that lead to unhealthy dietary habits is necessary for effective intervention programs.

Until now, large-scale dietary studies of food consumption used questionnaires and food diaries to keep track of the daily activities of their participants, 
which can be intrusive and expensive to conduct \cite{de1991seasonal}. 
Alternatively, social media is notorious for providing its users with a means of documenting the minutiae of their daily lives, 
including their dietary choices. 

\textbf{Can we use social media -- and Twitter in particular -- to get insights into dietary habits of an entire country?} After all, tweeting ``I'm having a sandwich'' has become a classic example of the ``pointless babble'' commonly found on Twitter.\footnote{A 2009 study found that 40\% of all tweets fell under the label of ``pointless babble'', see \url{http://www.pearanalytics.com/blog/2009/twitter-study-reveals-interesting-results-40-\ percent-pointless-babble/}. This study is feeding, pun intended, on a lot of such tweets.} Recent research into recipe search logs in US \cite{west2013cookies} and China \cite{zhu2013geography} were able to show temporal and spatial peculiarities of regional cuisines, suggesting that dietary habits may be closely linked to culture, as it is propagated geographically. However, such studies lack demographic and other personal information of their users, and are limited to coarse-grain geo-spacial analysis.
In this work we make a case for using social media to monitor dietary habits at both national and personal scale. 
We perform a large scale analysis of $210$K Twitter users in the United States, tracking their $502$M tweets. 
We augment this data using a variety of sources which allows us to consider the nutritional value of the foods mentioned in these tweets, 
demographic characteristics of the users who tweet them, their interests, and the social network induced by their interactions. 
We show that these food mentions reflect the state of national health by correlating them with the state-wide obesity and diabetes rates, finding a substantial correlation at $0.77$ and $0.66$, respectively. 
We also compare our food-based lexicon with previous CHI research \cite{Culotta14}, and show that our lexicon outperforms the generic LIWC one. Additionally, the single caloric value estimator (which does not require training) achieves the same performance as the model trained using 64 categories of LIWC.

\textbf{But is the data sensitive enough to pick out more personal dietary variations?} To answer this question, we use demographic information from the $2010$ US Census, extrapolate users' gender using their screen names, and characterize users' zip codes as being either urban or rural. We find gender differences among the Twitter users, with women generally tweeting about less caloric foods than men. 
Estimated education level also proves to be a significant factor, with fewer calories being mentioned in areas with higher education levels. Not only do we see a difference in the density of the foods mentioned by users, but we also find qualitative differences in their dietary selections. Alcoholic beverages tend to be mentioned in urban environments, whereas pizza and chocolate are popular in the rural ones. Users' self-disclosed interests are also related to their diets, with those who mention an interest in cooking decreasing the chance of being obese by $1.5$\%, which is in line with research showing that lack of cooking at home gives rise to obesity \cite{cutler13}.

\textbf{Does social network influence personal dietary habits of users?}   
Unlike traditional surveys that focus on isolated, randomly chosen individuals, social networks come with a \emph{network} structure. 
The links of this network allow to analyze individuals not just in isolation, but in the context of their social circle.
It has been suggested in \cite{gershenson2011epidemiology} that social trends can spread in society much like diseases, 
resulting in a \emph{social infection}. 
By inducing two kinds of social networks -- one using reciprocal communication and another using following relationships -- we show 
the assortativity in the dietary habits of users far beyond that which would be expected by chance.

Barring several limitations, which we outline in the Discussion section, we hope this study provides a case for the use of social media in public health monitoring in the dietary domain. We show that it is possible not only to detect indications of country-wide health trends, but zoom in on demographic and interest groups, potentially informing public health awareness campaigns.

\section{Related Work}
\label{sec:relatedwork}

Unlike in the animal kingdom, as omnivores, humans make nutritional selection 
not only based on their physiological needs, but also based on their culture and 
identity. As Fischler \cite{fischler1988food} puts it, by selecting and cooking 
food, one ``transfers nutritional raw materials from the state of Nature to the 
state of Culture''. The rules which are applied to food differ according to 
one's nationality, gender, and age, and social circumstances of the meal dictate 
its content, timing, and atmosphere. Below we outline the latest attempts to 
track public health using social media, and most notably nutritional research, 
which has been mostly thus far focused on recipe websites. 

Recently, Twitter has been used as a source of data for public health 
monitoring, such as for tracking flu-like symptoms 
\cite{aramaki2011twitter,culotta2013lightweight,sadilek2013modeling}, adverse 
side-effects of drugs \cite{bian2012towards}, tobacco use 
\cite{prier2011identifying}, and county-level health statistics \cite{Culotta14} . 
Using a text classifier, Sadilek \& Kautz 
\cite{sadilek2013modeling} detect tweets which mention the user being sick. They 
find that the higher social status of the users, the better their health, with 
poverty, education, and race (originating from the census data) explaining 8.7\% 
of the variation in observed health. The most predictive variables were 
proximity to polluted sites and encounters with sick individuals. 
Culotta \cite{Culotta14} uses well-defined lexicons such as LIWC and some demographic variables 
about users to predict county-wide health statistics (e.g., \textit{Obesity} and \textit{ Diabetes}) 
of the top 100 most populous counties in the US.
Prier et al.\ 
\cite{prier2011identifying} use LDA to find topics related to tobacco, such as 
addiction recovery, other drug use, and anti-smoking campaigns. Paul \& Dredze 
\cite{paul2011you} apply an Ailment Topic Aspect Model to tweets to discover 
mentions of various ailments, including allergies, obesity, and insomnia. More 
generally, life satisfaction has been mined from Twitter by Schwartz \& 
Eichstaedt et al.\ \cite{schwartz2013characterizing}, who evaluated their 
approach using phone survey data. Using LDA, they find word topics which 
correlate with demographics and socio-economic status, and provide insights into 
the sources of well-being, such as donating money and having rewarding jobs. 
These efforts, including those utilizing other social media websites like 
Craigslist \cite{haimson2014ddfseeks}, have aimed to augment the current data 
collection practices, making them faster, cheaper, and potentially more 
accurate.

In this paper we focus on the dietary choices of a large population of social 
media users. Culture-specific ingredient connections have been discovered by Ahn 
et al.\ \cite{ahn2011flavor} who mine recipes to create a ``flavor network''. 
Temporal nature of food consumption has been explored by West et al.
\cite{west2013cookies}, who mine logs of recipe-related queries. Using Fourier 
transforms, they illustrate the yearly and weekly periodicity in food density of 
the accessed recipes, with different trends in Southern and Northern 
hemispheres, suggesting a link between food selection and climate. Focusing on 
users who decided to go on a diet (as signified by them adding a book on dieting 
to their shopping cart), authors show the dip in caloric value per serving of 
the recipes users search for, and a gradual return to the pre-diet levels. 
Geographical distribution of food has been explored by Zhu et al.\ 
\cite{zhu2013geography}, who, unlike \cite{west2013cookies}, find climate 
(operationalized using temperature) to have little correlation with ingredient 
use, while finding geographical proximity to be a key factor in shaping regional 
cuisines. A recent work by Wagner et al.\ \cite{wagner2014spatial} on 
german-language recipe site shows similar negative relation between recipes and 
geographic distance of their seekers. We also find caloric content of foods 
mentioned in text a useful quantification of dietary selections, but unlike 
these studies, we relate it to public health statistics in order to validate its 
use. Furthermore, the nature of social media, unlike the recipe search logs, 
allows us to enrich our data with information on user demographics, interests 
and social network. In particular, our work on user interests echoes that in 
\cite{chunara2013assessing}, where interests of Facebook users were found to be 
related to their BMI. However, we take a more general approach to detecting user 
interests, using the network of the Twitter users.

Social nature of obesity has been hypothesized by Christakis \& Fowler 
\cite{christakis2007spread}, who tracked a densely interconnected social network 
of $12,067$ people across $32$ years. They found a person's chances of becoming 
obese increased by $57$\% if he or she had a friend who became obese in a given 
interval. These effects were not seen among neighbors in the immediate 
geographic location, emphasizing the importance of social ties. They provide 
three explanations for the collective dynamics of obesity: \emph{homophily} 
which is a tendency of people to associate with people who are similar to them, 
\emph{confounding} which occurs when people share attributes or jointly 
experience events, and \emph{induction} which refers to a person-to-person 
spread of behaviors and traits. These findings have been contested, however, by 
Cohen-Cole \& Fletcher \cite{cohencole2008is}, who claim ``social network 
effect'' becomes negligible once ``standard econometric techniques are 
implemented''. Recently, Silva et al.\ \cite{silva2014you}, who use Foursquare 
checkins to gauge food culture similarity between geographical locales, also 
found that countries closer in geographic proximity are not necessarily similar 
in their check-in behavior. In this work we examine both demographic and social 
aspects of food tweeting behavior, and provide some support to the social 
affinity that is not local in geographic sense.


\section{Data}
\label{sec:data}


We begin by collecting 50M tweets through the Twitter Streaming 
API\footnote{\url{https://dev.twitter.com/docs/streaming-apis}} using a keyword 
filter over a span of 2013/10/29 $-$ 2013/11/29. Keywords were selected to 
match 
as many food-related tweets as possible (covering eating, food, cooking and 
cuisine). The list of keywords contains also the names of the top $10$ 
fast-food 
brands in the US (e.g., McDonald's and Starbucks). 
Then we selected all geo-tagged tweets and filtered out those that are not 
posted from the US. The result was a collection of $892$K tweets posted by 
$400$K users from the US.
A uniform sample of $210$K users who contributed from the US was randomly 
generated. We requested for each user in the sample (through different Twitter 
APIs) the profile (i.e., name, description, location, \# tweets, \# followers, 
etc.), their latest $3.2$K tweets, and up to $5$K followers as well as $5$K 
friends. This process resulted in a collection of $503$M historical tweets, 
$44.5$M followers ($173$M links), and $32.1$M friends ($180$M links).    

We take several bootstrapping steps in order to improve our detection of 
food-related tweets and to extend our list of keywords. First, we examine the 
$1,000$ most frequent terms used in the tweets detected by the initial list, 
and 
hand-select other $118$ terms unambiguously related to food. This new ``food'' 
filter was now applied to the users' historical tweets. We then label a subset 
of tweets detected using these two filters, as well as a sample of those which 
were not thus far identified as food-related, in order to train a classifier. 
Using CrowdFlower\footnote{\url{https://crowdflower.com}} crowdsourcing 
platform, we published a total of $2,157$ tweet examples, collecting 3 
annotations for each tweet. The task was fairly easy, with $95.9$\% agreement. 
The training set, consisting of $583$ positive and $1,574$ negative examples, 
was used to train a unigram Naive Bayes classifier. 



Finally, we select $500$ 
most popular terms in the tweets the classifier deems to be on food-related 
topic and further enrich them with nutritional information -- mainly 
calories per serving. To estimate this value, we search a nutritional 
information website\footnote{\url{http://caloriecount.about.com/}} using the 
identified food keyword, such as \emph{pizza}. We then average the per-serving 
caloric values for the top $25$ returned entries (which in this case would be 
pizzas of different brands and with different toppings). We then manually check 
the validity of the resulting entries, excluding ambiguous ones such as 
\emph{plain} and \emph{cured}. The final list contains $460$ entries and is 
available online\footnote{\url{http://bit.ly/1whhowz}}.

In terms of food twitting activity, we detect weekly periodicity, as well as some major holidays: Thanksgiving, Fourth of July, and Valentine's Day. 
Weekly periods spike on Saturday and are at lowest on Mondays. 

Now, we can use this list to perform longest n-gram matching to detect the 
foods in the tweet text, and aggregate their caloric content. 
Upon manual examination of 800 tweets containing these foods, we found 70\% of tweets to mention some food, and out of these 63\% either explicitly or implicitly being about food consumption and another 12.5\% wishing for a food. Most false matches happened when the foods were used in figurative sense (``they scattered like fish'') or referring to others (``chocolate man''), though many of these references may still reveal cultural relationships with food, such as reference to Thanksgiving as ``turkey day''.
The most frequently 
mentioned foods in our dataset are \emph{pizza}, \emph{chocolate}, 
\emph{chicken}, \emph{ice cream}, and \emph{apple}. Among the top drinks are 
\emph{coffee}, \emph{beer}, \emph{wine}, and \emph{tea}. 
We then get the food that most distinguishes a given state, by first considering the 
top $200$ most 
popular foods in each state, and computing the difference in probability of each 
word from that of appearing in the overall food-related corpus -- a technique 
similar to feature selection for a binary classification -- to find the term 
most likely to be found in tweets of one state, and not the others. 
We find local peculiarities like California \emph{wine}, Florida \emph{orange}, 
Maryland \emph{crab}, and Alaskan \emph{salmon}. We also 
find possible erroneous matchings, as in the case of New York \emph{apple} 
(from 
``Big Apple'') and Missouri \emph{arrowhead}. These special cases, having 
particular prevalence in specific localities, attest the difficulty of 
identifying the proper context in a limited-length text of tweets. We leave 
further tuning of our food lexicon for future work.


\section{State-level correlations}
Although the cyclical nature of food consumption and major holidays can be 
detected in this data, there is still a concern whether the data sampling is 
representative of US population or, at least, useful to detect 
\emph{differences} in food consumption. Thus, we correlate the caloric values 
of 
foods mentioned in tweets per each state to the obesity 
rates\footnote{\url{http://www.cdc.gov/obesity/data/adult.html}} (from $2012$) 
and the incidence of 
diabetes\footnote{\url{http://www.cdc.gov/mmwr/preview/mmwrhtml/mm5743a2.htm}} 
($2005-2007$), as measured by Centers for Disease Control and Prevention (CDC). 
Obesity is defined by Body Mass Index (BMI) -- a person's mass divided by the 
square of their height -- with BMI $>$ $30$ considered obese. 



\newcommand\Tf{\rule{0pt}{2.8ex}} 
\newcommand\Ts{\rule{0pt}{2.2ex}}
\begin{table}
\caption{Pearson and Spearman correlations of tweet caloric value to 
state obesity and diabetes rates.}
\begin{center}
{\footnotesize
\begin{tabular}{l|cc|cc}\hline 
& \multicolumn{2}{c}{\textbf{Obesity}} & \multicolumn{2}{c}{\textbf{Diabetes}} 
\Tf\\\hline
 & Pearson & Spearman & Pearson & Spearman \Ts\\\hline
All            & $0.772^{***}$ & $0.784^{***}$ & $0.658^{***}$ & $0.657^{***}$ 
\Ts\\		
Food           & $0.629^{***}$ & $0.643^{***}$ & $0.538^{***}$ & $0.517^{**}$ 
\Ts\\		
Beverage       & $0.762^{***}$ & $0.786^{***}$ & $0.646^{***}$ & $0.622^{***}$ 
\Ts\\		
Alcoholic bev. & $0.445^{*}$ & $0.430^{*}$ & $0.073$ & $-0.007$\Ts\\\hline 	
\multicolumn{5}{r}{\footnotesize \textbf{Significance: $p<0.0001$ ***, $p<0.001$ 
**, $p<0.01$ *}\Tf}\\\hline
\end{tabular}}
\end{center}
\label{tbl:correlations}
\end{table}

Table~\ref{tbl:correlations} shows Pearson product-moment correlation $r$ and Spearman rank correlation 
coefficient $\rho$ between the average caloric density of tweets and these health 
statistics across the $50+1$ US states (including Washington DC). For each tweet, we use exact string 
matching to identify the foods (many of which would result in erroneous matches 
otherwise) and, if more than one is found, average their caloric value. In 
\emph{all} we consider all entries, we also differentiate between solid foods 
and non-alcoholic and alcoholic beverages. The correlation is the highest when 
we consider all foods, with the Pearson correlation of $0.772$ with obesity and 
$0.658$ with diabetes. For both ailments, beverage caloric value alone has 
higher correlations than solid food alone. However, the importance of alcoholic 
beverages differs drastically, with being somewhat correlated with obesity at 
$0.445$ and having no statistically significant relationship with diabetes. The 
reasons for this differentiation may be physiological, but also cultural. In 
the 
next section we also illustrate the association of alcohol with urban locales. 

\begin{figure}
\center
    \includegraphics[width=0.3\textwidth]{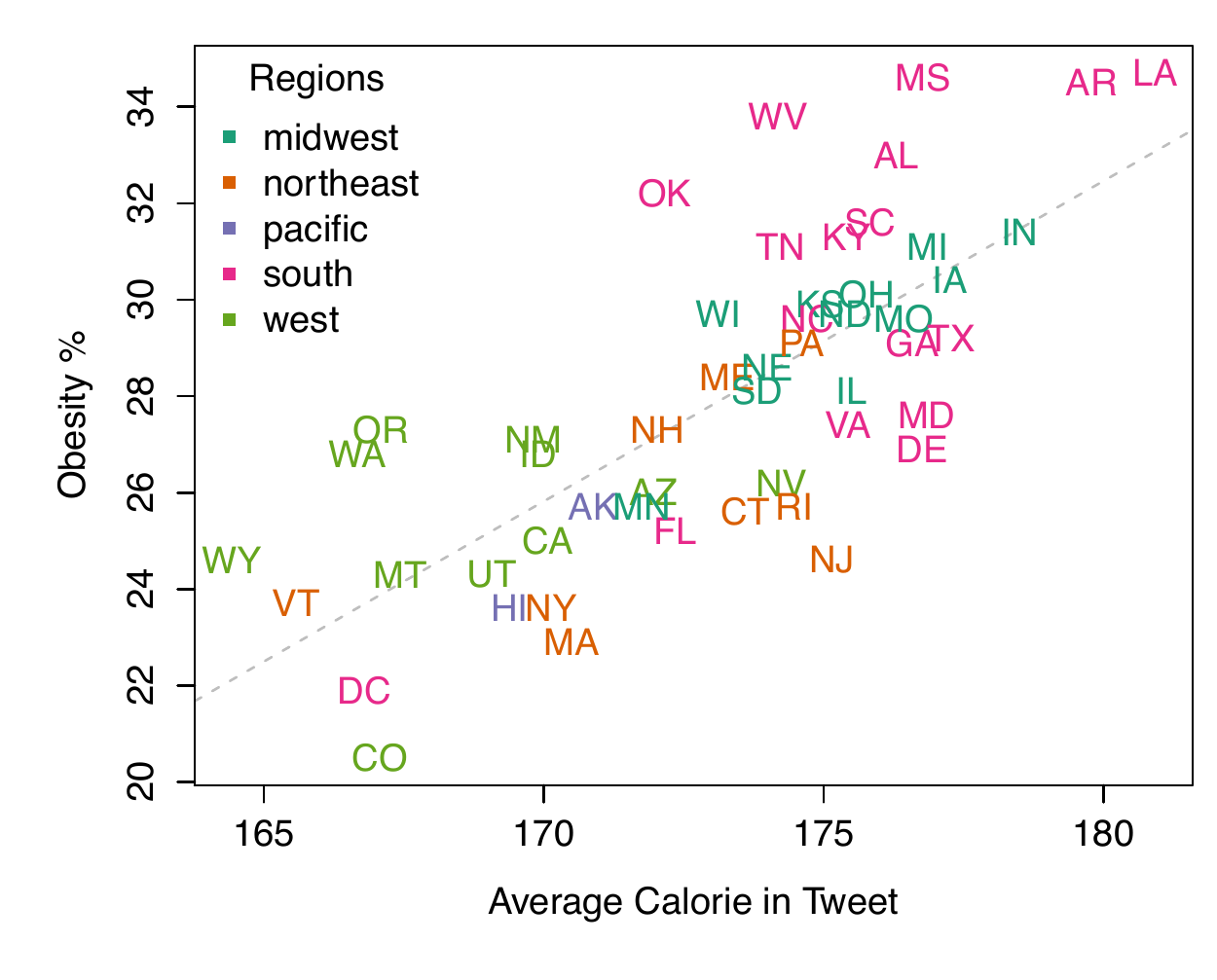}
\vspace{-0.3cm}
    \caption{Caloric value of foods mentioned in tweets versus obesity rates.}
    \label{fig:scatterplotstates}
\end{figure}

We further explore the relationship between obesity and caloric value of the 
mentioned foods in Figure~\ref{fig:scatterplotstates}, where we color the 
states 
according to their geographic region. The grey dashed line shows the linear 
regression line. We find the Southern states to be in the upper right corner, 
with Louisiana (LA) and Arkansas (AR) in the extreme right.\footnote{See 
\url{http://content.time.com/time/health/article/0,8599,1909406,00.html} on why 
that may be the case.} The clustering of the Southern, Midwest, and Northeast 
states suggests a common food culture between spatially proximal populations. 
These findings are supported by earlier work on recipe search in Germany 
\cite{wagner2014spatial} and ingredient use across China 
\cite{zhu2013geography}, who find that geographically closer cuisines shared 
more ingredients (with a few exceptions such as Hong Kong, which has a 
historical diversion from Chinese culture). Likewise, we notice Washington DC 
to 
be somewhat removed from its geographic neighbors, potentially due to the 
influence of the peculiar urban culture. We explore the effects of culture, 
personal interests, and demographics on the dietary habits of Twitter users in the next section.


\section{County-wide model fitting}
Despite the inherent sparsity problem observed at county-level as compared to the state level, we found
a Pearson correlation of the caloric value of all foods with obesity at $0.501$ and diabetes $0.447$ for
counties with at least $100$ users. For counties with at least $200$ users ($N=191$), the correlations were even better with $0.605$ for 
obesity and $0.498$ for diabetes.
Encouraged by this results, we wanted to investigate further the extent to which food mentions could capture  
county-wide health signals such as obesity and diabetes. To this end, we designed an experiment that compares
our ``food mention'' model to the one presented in \cite{Culotta14}. 
Similarly to the paper we train a regression model using different kinds of variables to predict obesity and diabetes rates at county level. 

Culotta's model \cite{Culotta14} uses lexicon categories along with a selection of demographic variables
to predict county-level health statistics of the top 100 most populous counties in the US.
Their experimental study revealed a strong predictive accuracy of demographic variables which can 
be improved if combined with linguistic variables derived from LIWC\footnote{\url{http://www.liwc.net/index.php}}
(Linguistic Inquiry and Word Count) lexicon.

To enable the comparison with Culotta's model, we first build a user model with the following variables:
\begin{itemize}
\item LIWC categories: For each user, we compute a binary vector of LIWC categories extracted from her 
Twitter profile. A LIWC category (e.g., \textit{Social}, \textit{Family}) takes the value of 1 if 
the user mentions at least one word in her profile that belongs to the category, 0 otherwise.
\item Food names: For each user, we compute a binary vector of Food names mentioned in her tweets. 
We use our handcrafted dictionary of food names and apply an exact matching to the tweets.  
\item avgCal: this variable reflects the average caloric value computed across all food names mentioned
by the user in her tweets. 
\item Demographic variables: Each user is assigned a list of five demographic variables derived
from census data related to the county to which she belongs. These variables are: \textit{Under\_18} 
(proportion of people under the age of 18), \textit{Over\_65} (proportion of people above the age of 65)
, \textit{Female} (proportion of females), \textit{Afro-Hispanic} (proportion of Afro-American
and Hispani), and \textit{Income} (log of median household annual income). 
\end{itemize}

Next, we aggregate user models  at the level of counties. 
We generate for each county, a vector of LIWC categories (64), food names (461), and demographic
variables (5). The weight of a given food name (resp. LIWC category) reflects the proportion of users in that
county who mentioned that food name (resp. LIWC category). 
Note that unlike Culotta's work where only the top 100 most populous counties (based
on census data) are considered, we have retained all counties with at least 100 users (346 counties).

We run a series of regressions using various models (features) to predict obesity and diabetes scores of the 346 counties.
We consider six models: \textbf{Demog} (demographic variables), \textbf{Liwc} (LIWC categories), \textbf{Calories} ($avgCal$ variable), 
\textbf{Food} (food name variables), \textbf{Liwc-Demog} (LIWC categories and demographic variables), and \textbf{Food-Demog} 
(food names and demographic variables).  
For each model, we use a five fold cross-validation to assess the generalization of its accuracy.
Folds are selected in a way that prevents counties from the same state to appead simultaneousely in both training and test sets. 
Finally, we use Ridge regression in order to reduce overfitting, espacially in models with large number of variables (e.g., Food (461)).

\begin{figure}
    \includegraphics[width=0.42\textwidth]{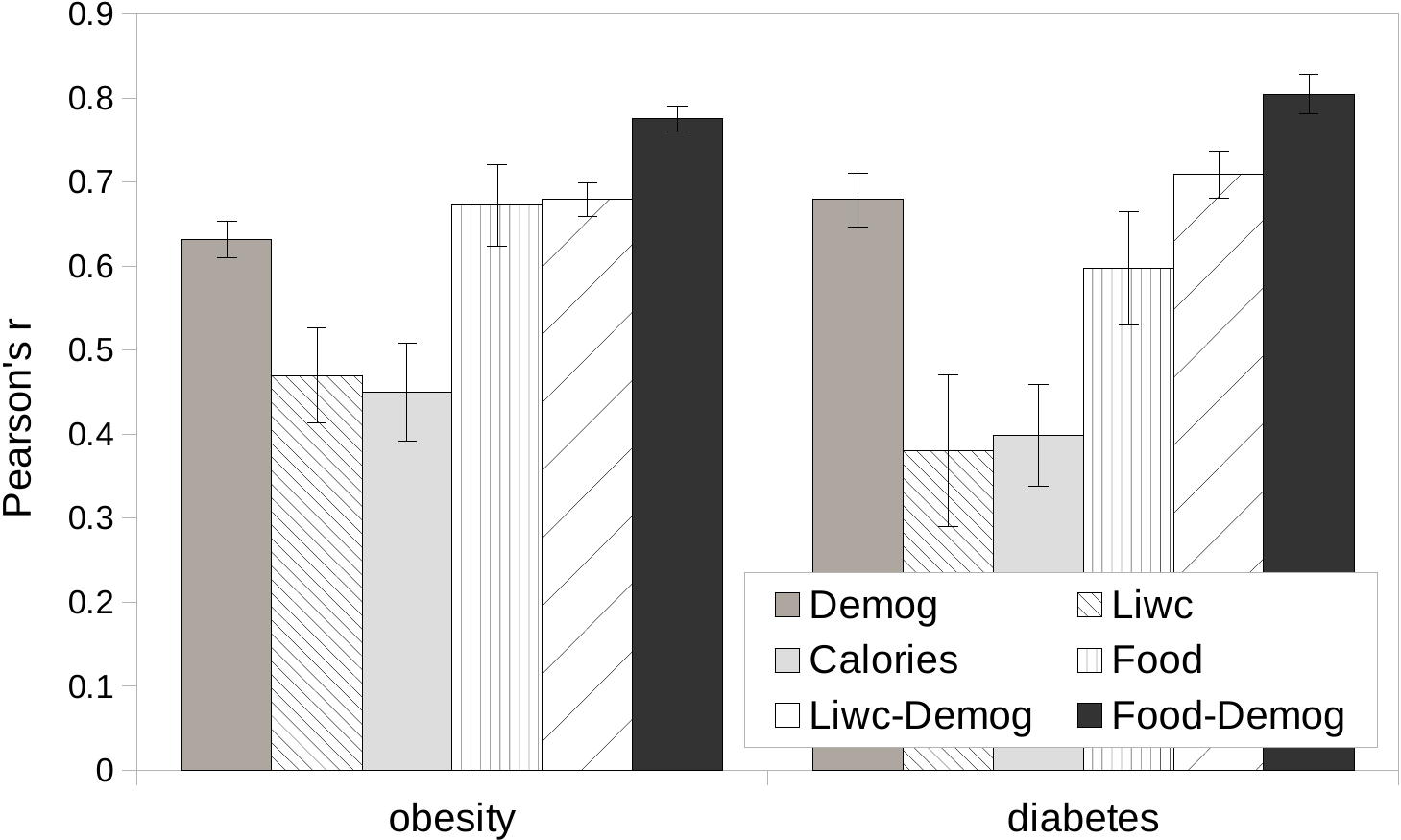}
    \caption{Held-out Pearson's correlation with standard error of the mean (SEM) for 
the different models. Demog: uses the five demographic variables. Liwc: consists of 64 LIWC categories,
Calories: consider only one variable avgCal. Food: consists of 461 food names. 
Liwc-Demog: combines LIWC categories and demographic variables. Food-Demog: combines food names and 
demographic variables.}
    \label{fig:liwc}
\end{figure}

Figure \ref{fig:liwc} shows the Pearson's correlations (which are all statistically significant) achieved by different
models along with their corresponding SEM (standard error of the mean) scores. Recall that the SEM score 
is equal to the corrected standard deviation of the sample divided by the root square of the size of that sample. 
\textbf{Food-Demog} model, which combines food names and demographic variables, is found to outperform
Culotta's model \textbf{Liwc-Demog} which combines LIWC linguistic variables with 
Census demographics. If fact, \textbf{Food-Demog} achieves held-out correlations of 
0.775 for obesity and 0.804 for diabetes while \textbf{Liwc-Demog} achieves respectively 0.679 and 
0.708 for the two health statistics. 
Surprisingly, the simple \textbf{Food} model based on our handcrafted dictionary of food names 
significantly outperforms the \textbf{Liwc} model. 
Also, we notice that \textbf{Calories} model which has only one variable achieves reasonably good correlation scores
(0.450 for obesity and 0.398 for diabetes) compared to \textbf{Liwc} 
(0.470 for obesity, 0.380 for diabetes) which has 64 variables. 
Note that we also tried a baseline using top 1,000 most frequent hashtags in our dataset and find very low correlations with the county-wide obesity and diabetes rates of around 0.06. Finally, we built another baseline that uses generic tweet statistics, including number of tweets, retweets, number of replies, and number of hashtags. This baseline achieved comparable correlations (0.38 for obesity and 0.44 for diabetes) to those of the ``Calories'' model. 
\note[Yelena]{Discuss why this generic model may be performing so well.}

Finally, we observe that \textbf{Food} model significantly outperforms -- the aggregated -- \textbf{Calories} model for the prediction of 
both obesity and diabetes.
Yet, \textbf{Calories} model has several advantages. 
First, \textbf{Calories} model is much simpler than all the other models as it relies on a single variable ($avgCal$) and does not require 
any model fitting. 
Second, the model is easily interpretable and arguably closer to the ``root cause'' of obesity and diabetes. 
Third, the caloric value of food has shown very strong correlations at the state level. 
Next, we use \textbf{Calories} model and the best-performing model \textbf{Food-Demog} to zoom in to the personal, user level of analysis.

\section{Characterizing Users}
\label{sec:users}

\subsection{Income, Education and Gender}

The effect of obesity in US has been shown to vary according to income, education, and gender. According to CDC, whereas among men, obesity prevalence is generally similar at all income levels, higher income women are less likely to be obese than their low income counterparts\footnote{\url{http://www.cdc.gov/nchs/data/databriefs/db50.pdf}}. Motivated by these statistics, we  map the income and education figures from 2010 US Census\footnote{\url{www.census.gov/2010census/data/}} to zip codes from which Twitter users sent their messages. As the statistics are broken down into bins (such as ``\$$10,000$-\$$14,999$'' for income and ``Some College'' for education), we computed a weighted average of these values using the mean of each bin. Finally, we associated each user with the zipcode most frequently associated with their tweets. 

We supplement this with a per-user gender classification using Genderize API\footnote{\url{http://genderize.io/}} on users' screen names. The API uses a database of names from major social networks, and produces the most likely gender associated with a first name, which can be \emph{male}, \emph{female}, or \emph{none} when the gender cannot be detected. Among our users we detect $37.2\%$ as female, $32.1\%$ as male, and the remaining $30.7\%$ were labeled as none. Concerned about the latter group, we run a crowdsourced experiment on CrowdFlower to tag a random subset of $1,331$ accounts as belonging to a real person or not. Whereas the female and male users had low numbers of non-personal accounts ($2.57$\% and $3.49$\% respectively), over a quarter of ``none'' gender -- $26.69$\% -- were such accounts. For this reason, we exclude these users from further analysis, with $128,487$ data points remaining. Notice that we did not remove users with ``none'' gender from the previous computation of correlations because of their limited impact on the results. For instance, removing these users brings Pearson correlation between caloric values of all food and obesity from $0.772$ (all users) to $0.741$.

\begin{figure}
    \begin{center}
    \includegraphics[width=0.47\textwidth]{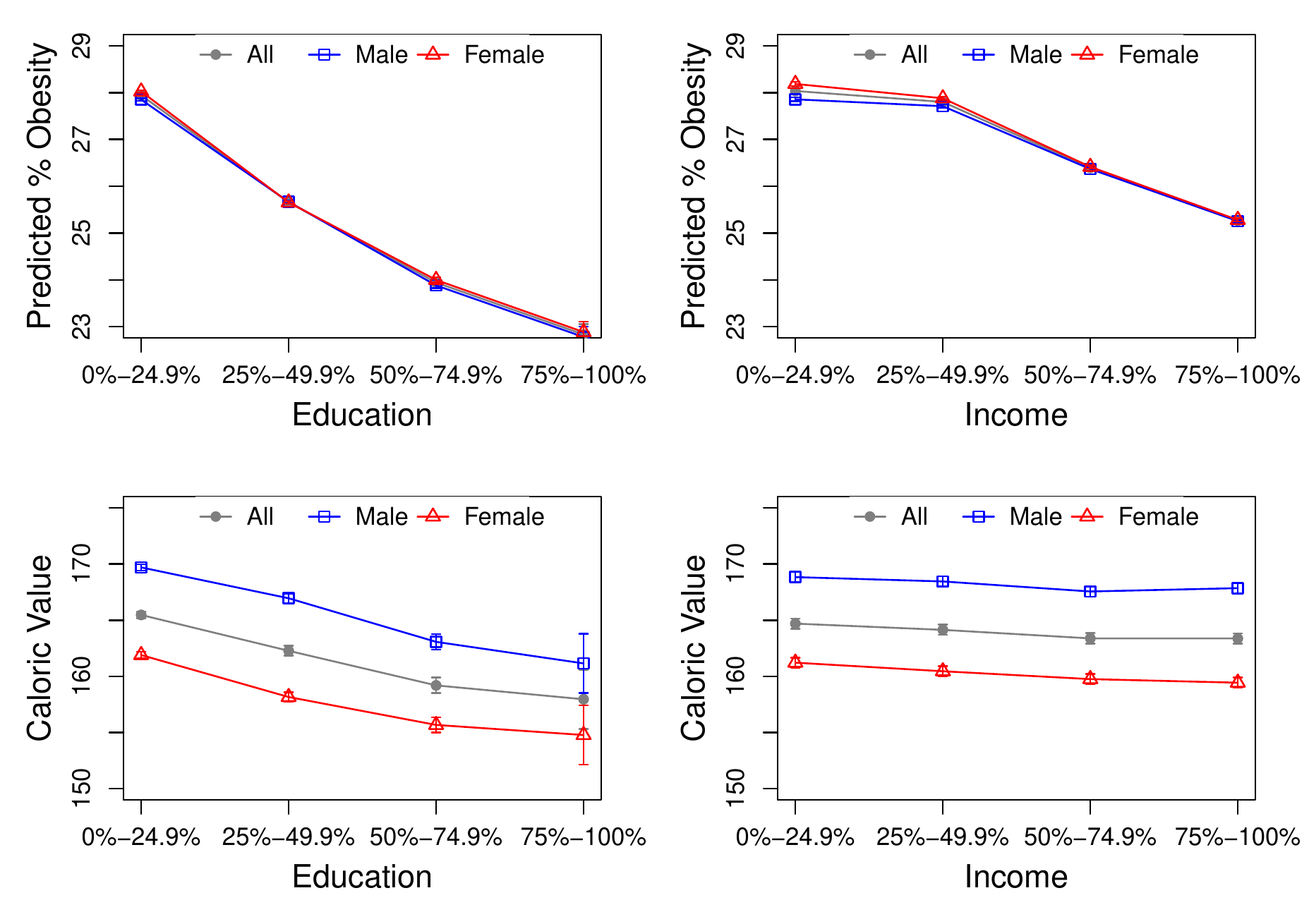}
    \caption{Average predicted obesity and caloric value of tweets of four quartiles divided by education (Bachelor degree attainment, and income level), with 95\% CI.}
    \label{fig:eduincomegender}
		\end{center}
\end{figure}

Figures \ref{fig:eduincomegender} shows an exploration of the dimensions of gender, educational attainment (likelihood of a Bachelor's degree) and median household income. Two variables are plotted: the likelihood of obesity (estimated using the model built in the previous section) the average caloric values of the tweets produced by each user. We find that although the trends differ in for the genders in the kind of heaviness of food they tweet, the distinction is not present in the obesity rates. It is understandable, given gender-mixed populations on which the model was trained. We witness a stronger negative correlation for educational level with both the caloric values and the estimated obesity rates. These trends echo those found by USDA, which finds obesity prevalence increase as education decreases (though more so for women).



Next, we mapped the zip codes to major metropolitan areas (top $100$ US cities by population) and labeled all of these as ``urban''. We find $36,196$ ($26.9\%$ of all) users to be in these urban areas. Although one may suppose eating at restaurants (in urban areas) may result in tweets mentioning more caloric food, we find quite the opposite. There is a significant (at $p<0.001$) distinction between the two populations, with users in rural areas having an average of $164.8$ calories in their tweets, and urban $161.6$.

\begin{figure}[h]
\centering
\begin{subfigure}{.23\textwidth}
  \centering
  \includegraphics[width=\textwidth]{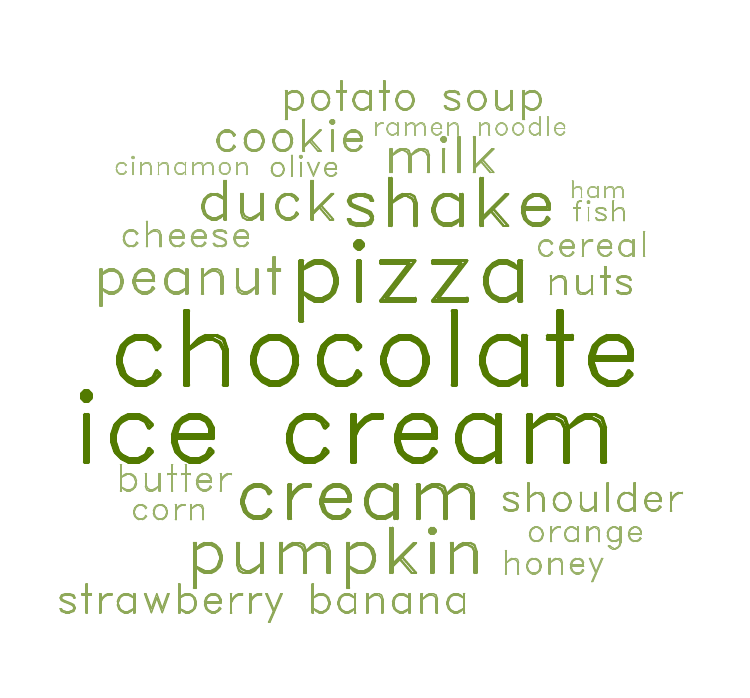}
  \vspace{-0.7cm}
  \caption{Rural}
  \label{fig:wordcloudRural}
\end{subfigure}%
\begin{subfigure}{.23\textwidth}
  \centering
  \includegraphics[width=\textwidth]{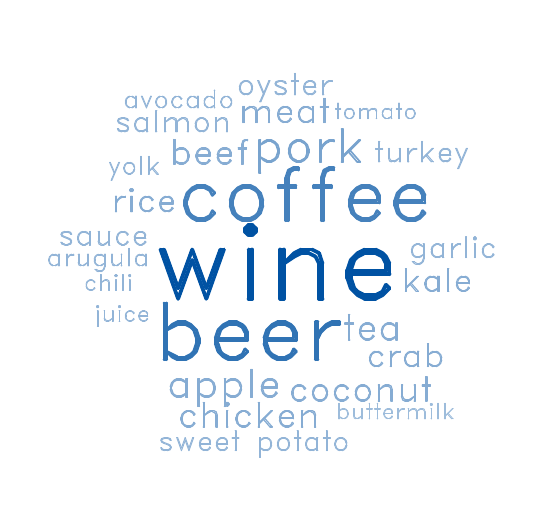}
  \vspace{-0.7cm}
  \caption{Urban}
  \label{fig:wordcloudUrban}
\end{subfigure}
\caption{Distinguishing foods between rural and urban users.}
\label{fig:wordclouds}
\end{figure}

The above distinctions are identifiable not only in the caloric value of the foods mentioned in the tweets, but in the foods themselves. Figure~\ref{fig:wordclouds} shows the distinguishing foods between users in rural and urban areas, computed using probability differences such that the foods which are popular in one area but not in another get a higher score. Urban food vocabulary distinguishes itself with alcoholic drinks (\emph{wine} and \emph{beer}), and more uncommon foods like \emph{avocado} and \emph{crab}, whereas the rural cuisine emphasizes \emph{pizza} and common deserts like \emph{chocolate} and \emph{ice cream}.

\subsection{Interests}

There are, of course, other factors which contribute to one's diet. For instance, people who are interested in healthy eating and organic food, or those with families would consciously alter their diets. To detect individuals with such interests, we compiled a high-precision keyword lists to detect users interested in, or at least mentioning cooking, dieting, organic food, and health. We also detected users who mentioned being a family member (being a father, mother, having kids, etc.). These filters were then applied to user profiles. 

The above method can be used to detect interests users feel comfortable declaring in their profile, however not all user characteristics can be extracted. We take a different approach to detect users who declare themselves to be overweight by the use of hash tags such as \texttt{\#fatgirlproblems} or \texttt{\#fatguyproblems}. These are typically used by users in self-reference, for example ``\emph{I have more pictures of food than I do selfies .. \#fatgirlproblems}'' Doing this, we detected $10,797$ users using at least one such tag. 

Table~\ref{tbl:interestcorrels} shows the difference in estimated obesity of population for which a variable has been detected (say, mentioning sports) from that where the variable was not detected, as well as the number of instances found for the class. Thus, if the difference is positive, the class has a higher chance of obesity by that many percentage points. Note that for this experiment only users whose names resulted in an identified gender were used (totaling in $128,487$ data points). We find that people who are interested in cooking and organic food decrease their likelihood of obesity by $1.3$ and $2.1$ percentage points, respectively. Although our \emph{on diet} keyword filter produced a low recall -- of only $107$ users -- the chance of decrease is near $1$\%. The detection of \texttt{\#fatproblems} hashtags resulted in $10,797$ matches, but we see only a moderate (although statistically significant) positive change in the obesity rate. 

\begin{table}[t]
\parbox{.98\linewidth}{
\caption{Difference between the estimated obesity rate while considering profile factors (mean rate at factor = 1 minus mean rate at factor = 0), and \# of non-zero instances. }
\begin{center}
{\footnotesize
\begin{tabular}{lrrrr}\hline 
\textbf{factor}     & \textbf{$f_1$ - $f_0$} & $p$-value & \textbf{non-zero $n$}\Tf\\\hline

\texttt{\#fatproblems} & +0.3 & 0.0000 & 10797\Tf\\
Student       &  0.0 & 0.3695 & 2709 \\
Family        & +0.3 & 0.0000 & 8014 \\
Cooking       & -1.3 & 0.0000 & 637 \\
Organic       & -2.1 & 0.0005 & 49 \\
On Diet       & -0.9 & 0.0174 & 107 \\
Health/Sport  & -0.2 & 0.0138 & 3181 \\\hline
\end{tabular}}
\end{center}
\label{tbl:interestcorrels}

\vspace{0.5cm}
\caption{Top 15 factors by the magnitude of the coefficient in a linear regression modeling the predicted obesity rate using interests as determined by following WeFollow users. For binary variables, the number of non-zero instances is given. }
\begin{center}
{\footnotesize
\begin{tabular}{lrrr}\hline
\textbf{Factor} & \textbf{Coefficient} & \textbf{Sign.} & \textbf{non-zero $n$} \Tf\\\hline
(Intercept) & 26.98593 & *** &  \Tf\\
tvshows & 0.80296 & *** & 4267\Tf\\
education & 0.63017 & *** & 1904 \\
business & 0.62499 & *** & 1962 \\
sports & 0.49467 & *** & 11423 \\
nfl & 0.35523 & *** & 6111 \\
entrepreneur & -0.23009 & ** & 5530 \\
music & -0.28651 & *** & 20364 \\
baseball & -0.29567 & ** & 1990 \\
animals & -0.29867 & ** & 1510 \\
travel & -0.36032 & *** & 3220 \\
news & -0.47807 & *** & 9266 \\
blogger & -0.68136 & *** & 7209 \\
football & -0.92379 & *** & 4691 \\
media & -1.08349 & *** & 6782 \\
tech & -1.15935 & *** & 3804 \\ \hline
\multicolumn{4}{r}{\footnotesize \textbf{$p<0.0001$ ***, $p<0.001$ **, $p<0.01$ *}} \Tf\\\hline
\end{tabular}}
\end{center}
\label{tbl:wefollowregression}
}
\end{table}

We go further and identify fine-grained interests as in \cite{abisheva2013watches}. Using WeFollow\footnote{\url{http://wefollow.com/}}, we collect a list of users who are judged to be prominent in some area. For each of the 61 areas such as \emph{TV}, \emph{Science}, or \emph{Football}, we collected top $200$ users, having the highest prominence score (ranging from $0$ to $100$) in their area. We compute a user's aggregate interest score for an area of interest by summing over the scores of the prominent users they follow. However, following behavior differs among the various areas. For example, an average aggregate score for \emph{Social Media} is $311$, whereas that for \emph{Cats} is only $64$. Thus, we consider a user to have an interest in an area if the aggregate prominence score of her friends in that area is at or above the mean score of all users potentially interested in that area. Using these variables we construct a linear regression model. Table \ref{tbl:wefollowregression} shows the top significant factors by the magnitude of the coefficient. Among these we see interest in TV shows and general sports categories to have positive relationship with obesity, whereas interest in football (soccer) and technology has the opposite effect. Recall that these interest scores do not stem from the text of the tweets, but from the user's following network. In that way, they provide a glimpse of a users' interests which may be not be available in their tweets, but which still may provide some indication of the user's dietary health.

Our findings partially confirm a previous study on Facebook interests of users in US and New York metropolitan area \cite{chunara2013assessing} in which it was found that a greater proportion of the population with interest in television was associated with higher prevalence of obesity. However, our observations of interests in sports is less straightforward, with \emph{sports} and \emph{NFL} categories positively relating to obesity, whereas the Facebook study finds activity-related interests to be associated associated with a lower predicted prevalence of obese and/or overweight people. The separation between watching and participating in sports would shed more light on this discrepancy.

\section{Social Nature of Food}
\label{sec:social}

Social circumstances play an important role in how we consume our food, and which food we consume. We attempt to discern the relationship between social interactions,
expressed in the text of the tweet and in the follower network of the user, and users' eating behavior 
(operationalized by predicted probabilities of obesity and diabetes, frequency of food mentioning, and caloric value of tweets).
We first study the impact of two types of relationship networks on the obesity and diabetes scores of users. 
Then, we use a threshold model to quantify the circumstances under which a user would get exposed to higher obesity and diabetes 
 risks. Finally, we look at the influence of friends at varying degrees of closeness.

\subsection{User-level obesity, diabetes and food frequency}
In this section, we focus on the impact of social relationships on obesity, diabetes and the amount of food related tweets. 
For each user, we compute the fraction of tweets citing at least one food name. 
To estimate the obesity and diabetes scores at individual levels, we use the \textbf{Food} model that relies on food names only.
For training a model, each user is assigned the obesity and diabetes rates of their county. A Ridge regression is used to learn
the models and predict individual user scores. 
Predicted scores could be seen as a user's risk level of obesity and diabetes, based on the food mentioned in their tweets.

Note that for the following analysis we deliberately discarded demographic variables. Friends on Twitter are also likely to live in close proximity \cite{takhteyev2012geography} and hence are likely to share similar estimates of demographic variables, based on geographic census data. So friends are likely to be similar along this dimension. However, we are more interested in observing the connection between social network closeness and propensity to share ``unhealthy tweeting behavior'', such as the use of food names associated with higher obesity levels.

\subsection{User Networks}
We explore the social nature of eating habits by constructing two social networks, namely \textit{Friendship network} (FN) 
and \textit{Mention network} (MN).
The Friendship network relates to the structural aspect of Twitter. In FN, two users are considered to be ``friends'' only if they follow each other. 
This definition follows the principal of mutual reachability introduced by 
Xie et al.\ \cite{Xie:2012} to identify real-life friends from Twitter. 
Alternatively, Mention network relates to the behavioral aspect of users. 
Here, a user $a$ (mentioner) has a link to user $b$ (mentionee) only if $a$ has mentioned $b$ in at least one tweet. 

Note that users without gender have been removed from both MN and FN, and only links between users with a known gender 
(i.e.\ \textit{male} or \textit{female}) in the initial set of US users are considered. 
Table~\ref{tbl:networks} provides some statistics on the two networks. 

\begin{table}[ht]
\caption{Description of Friendship and Mention networks}
\begin{center}
{\footnotesize
\begin{tabular}{c|cc}\hline
\textbf{Network} & \textbf{\# Links} & \textbf{\# Users}  \Tf\\ \hline
Friend  \Tf& $295,285$ & $84,599$ \\ 
Mention & $378,801$ & $85,144$ \\ \hline
\end{tabular}
}
\end{center}
\label{tbl:networks}
\end{table}

\subsection{Obesity and Diabetes Activation (Spread)}
Focusing on obesity, we ask to which extend does being connected to users with a high likelihood of obesity and diabetes (based on the food names they tweet) increases the likelihood of a  user
for these two health issues? 
A threshold model postulates that a success or failure of a social diffusion process depends on the reaching of a 
certain critical number of adopters \cite{valente1996social}. 
This model has been used, for example, to model the adoption of a 
particular notation for source attribution on Twitter \cite{kooti2012predicting}.
Following the idea of ``social activation'', we label users beyond the $90^{th}$ percentile (i.e.\ the top 10\%) in terms of obesity and/or diabetes scores as ``\textit{active}'' users. 
Then, for every user we calculate the number of active users to whom they are connected in both friendship and mention networks.
Finally, we compute the activation probability (i.e., the probability of being an \textit{active} user) 
given that a user is connected to $x$ \textit{active} users.

\begin{figure}[t]
\centering
\begin{subfigure}{.24\textwidth}\includegraphics[width=\textwidth]{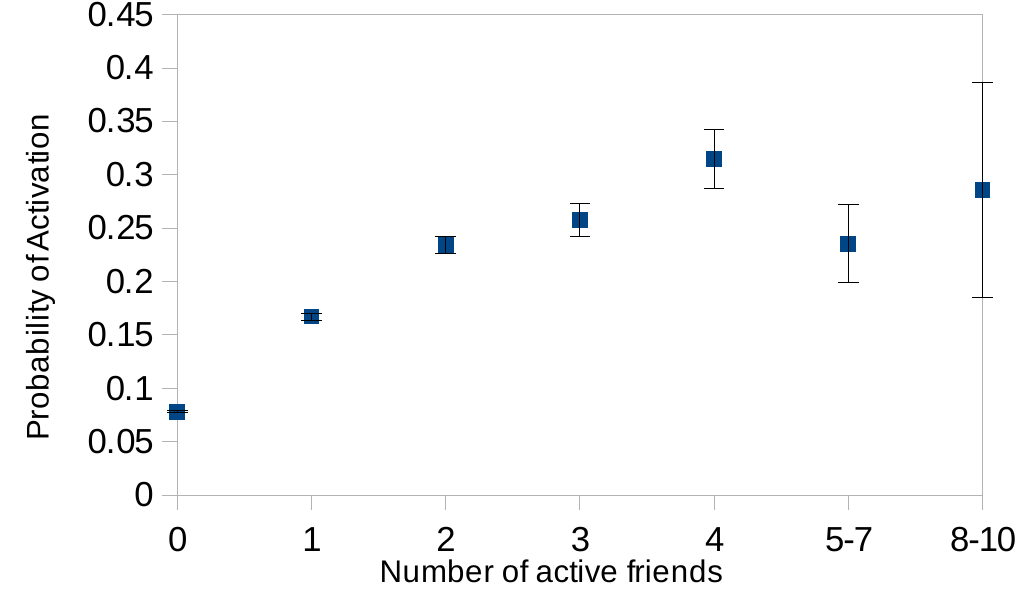}\caption{Obesity spread in FN}\label{fig:fn_obesity}\end{subfigure}%
\begin{subfigure}{.24\textwidth}\includegraphics[width=\textwidth]{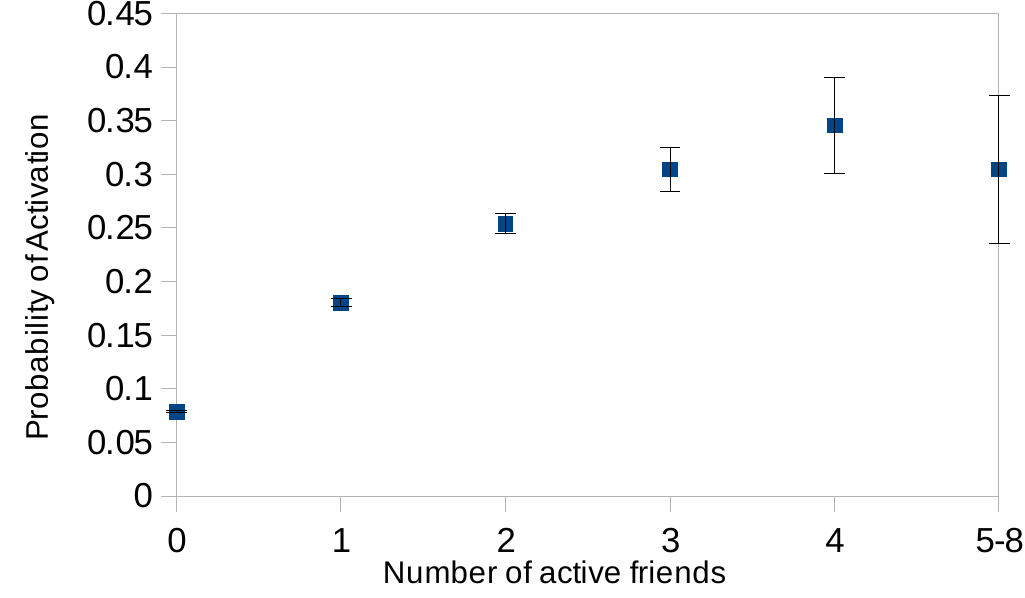}\caption{Obesity spread in MN}\label{fig:mn_obesity}\end{subfigure}%
\caption{Obesity activation probabilities function of the number of ``\textit{active}'' friends 
in Friendship Network (FN) and Mention Network (MN).}
\label{fig:activation}
\end{figure}

Figure~\ref{fig:activation} shows the activation probability scores of users
as a function of the number of their \textit{active} friends. 
As expected, the probability of showing a strongly increased probability of obesity, estimated by the food names mentioned, increases as the number of friends with high obesity scores increases. This increasing trend is particularly pronounced for up to four active friends. Though the standard errors of the computed probabilities increase dramatically beyond this point, 
with only few users having more than five active friends, there seems to be a plateau effect. The same behaviour is observed in the activation graph of diabetes.  

To alleviate the influence of \textit{content spread phenomena} in which people may tweet about a food name just because their friends tweeted about it, we removed all tweets that are ''replies to'' or ''retweets of'' other tweets. We trained a new \textbf{Food} model and run our activation algorithm which resulted in the same activation curve as Figure~\ref{fig:activation}. 
Thus, although we cannot exclude offline motivational effects, excluding them from data does not affect our results.
We also investigate the effect of \textit{geography} to make sure that the activation process is not the result of people living in the same region as their friends. We removed from the social graph all friendship links between users living in the same US state. We observed a small drop in activation probability scores, yet the increasing trend function of the number of active friends was still there. 

\subsection{Clique-ness Analysis}
Another way to assess the influence of social connections on users' dietary habits is to take into consideration the strength of 
user connections. Here, we define the link strength between any pair of users $(a,b)$ having an edge in the mention network or 
the friendship network as the 
fraction of their common friends in the same network. 
As mentioned in the dataset description (section Data), we have collected up to $5,000$ friends 
for each of the $210$K US users, leading to a user-friend bipartite graph with $\approx$180M links. 
Then, we use Jaccard index to 
compute the similarity of the friend sets between all pairs of users in both networks. 
Based on the computed scores, users and their links are 
assigned to different bins corresponding to different Jaccard index intervals. 
As expected, the distribution of Jaccard scores is 
heavily skewed. For instance, more than 71.6\% of the total number of links have a strength score lower that $0.1$ in the mention network.

 
\begin{figure}[h]
\begin{center}
\includegraphics[width=0.39\textwidth]{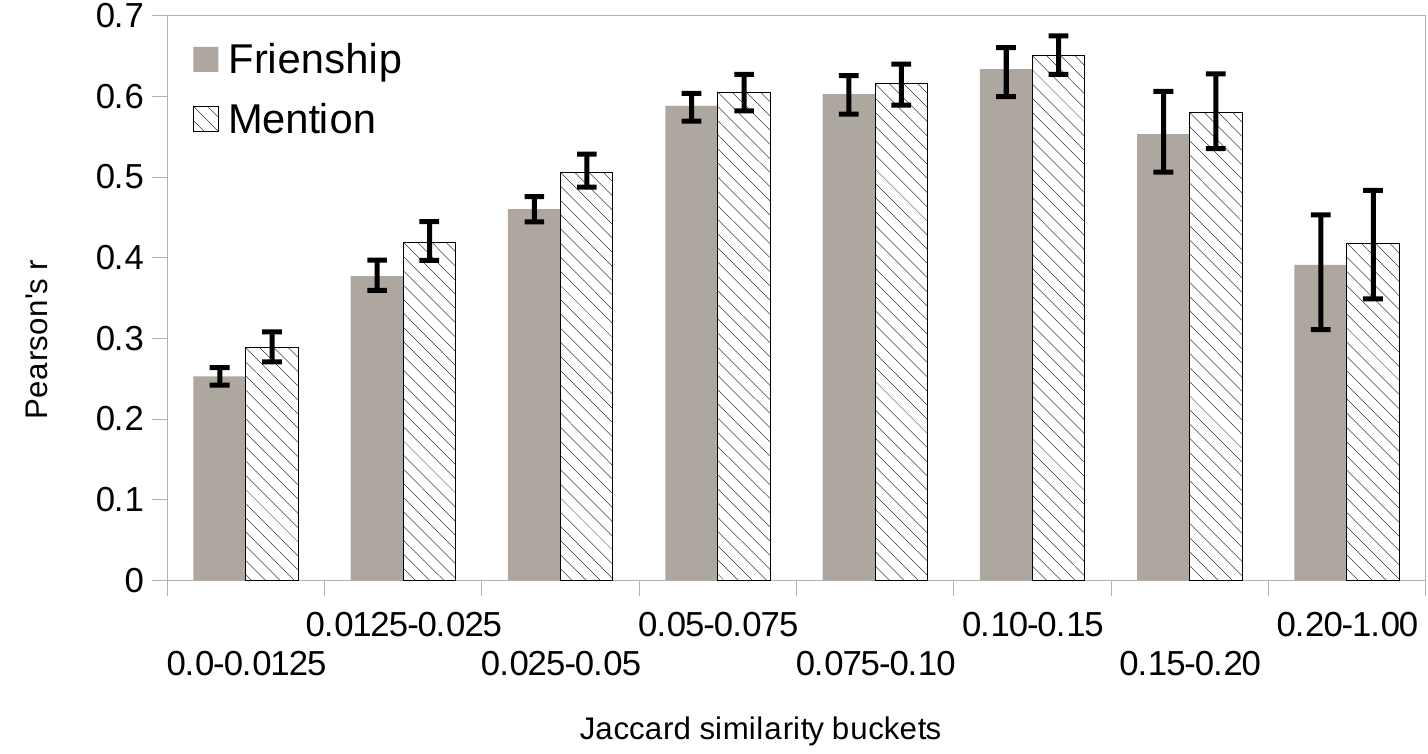}
\caption{Pearson correlation of the fraction of food related tweets between users and their friends, bucketed by Jaccard similarity 
in mentions and friendship networks. The 95\% confidence intervals (CIs) are computed using a bootstrap test, iterated 1,000 times.
CIs correspond to the $2.5^{th}$ and $97.5^{th}$ percentiles obtained over the bootstrapping process.}
\vspace{0.5cm}
\label{fig:jaccard}
\end{center}
\end{figure}

In order to check whether tightly connected friends share a higher degree of content similarity, we correlate the fraction of food 
tweets of each user to that of their friends within each bin. Figure~\ref{fig:jaccard} shows the bins from most dissimilar 
users ($[0.0,0.0125]$) to those most similar in terms of ego-network overlap. Note that the intervals are not normalized due to the highly skewed distribution 
of the data, with the last bin of $[0.2,1.0]$ having only $5,428$ users in Mentions and $5,076$ in Friendship networks. 
We witness an increase in the correlation of the food-related tweet fraction of users and their friends as network overlap 
increases until the last two buckets -- those with users sharing the most friends. 

Upon manual examination, we find highly related users to be in the same social locales, such as students in the same university or school. 
The phenomena may also be related to the shape of the \emph{exposure curve}, as proposed by Romero et al.\ \cite{romero2011differences}, 
which models the rise of initial interest as the number of exposures increases, and the decrease after over-exposure to a phenomenon.



\section{Discussion \& Future Work}
\label{sec:discussion}

Using data from social media to study sociological phenomena and to ``nowcast'' variables such as obesity rates has both advantages and disadvantages compared to traditional survey-based methods. Two of the most commonly given advantages include cost and timeliness. Data for hundreds of thousands of users can be obtained at relatively low cost, and this data can be collected, aggregated and analyzed in a matter of days or even hours, rather than weeks or months for comparable surveys. However, in addition to these two quantitative differences, there is also an important qualititative advantage: the availability of network information. Having access to even a subset of people that a given person interacts with makes it possible to look for evidence of social mechanisms such as homophily. In the health domain, having data on the strength of such effects makes it possible to consider \emph{social} interventions. Rather than trying to change unhealthy behavior of individuals by providing information and incentives derived solely from their own behavior, information about their friends' physical activity or healthy dietary choices could be provided. Similarly, altruistic incentives of the type ``Run a total of 30km this week and your friend @JohnDoe has a chance to win an iPhone'' could be based on automatically derived, intricate knowledge of a user's social circle. Or if a set of friends are planning a physical activity, they may be encouraged to invite others from their social circle.

The availability of rich data related to hobbies and interests is also a potential advantage for studies using social media. Public health initiatives targeting various segments of population already exist, such as UK's Change 4 Life (\url{http://www.nhs.uk/change4life}) targeting parents and USDA's MyPlate on Campus for university students (\url{http://www.choosemyplate.gov/MyPlateOnCampus}). In our study, we link users' obesity likelihood, inferred by a model combining demographic estimates with food names mentioned on Twitter, to their interests, such as TV shows, sports, and movies. The relationship between each of these activities and food would help target population segments which are especially at risk, and maximize the return on media advertising expenditures. But it is our vision that automated tools will help provide personalized messages to social media users, providing context-aware, real-time information, suggestions, and motivation. For instance, if a user tweets about an intention to exercise, links to tutorials and videos can be suggested. The development of these approaches first requires automated assessment of a user's dietary behavior, and this paper takes first steps in that direction.

Of course, sociological studies which use social media also have significant drawbacks. Typically, such studies suffer from a user sampling bias with an over-representation of affluent and tech-savvy demographic groups. This bias can be more pronounced when only users with GPS-enabled devices are considered.  
Indeed, the users in our subset come from neighborhoods with average household income at $85,117$, well above the US average of $51,017$ in $2012$\footnote{\url{http://www.census.gov/prod/2013pubs/p60-245.pdf}}. 
Similarly, our users come from locations where the average percentage of people with a Bachelor degree or above is at $23.71 \%$, slightly higher than the nationwide percentage of $22.23 \%$ observed in Census database. 
This, however, may be an artifact of the aggregation of census statistics (which are binned for each district in a range with no other distributional information), and a more fine-grained analysis may improve these numbers. Still, our sample did closely resemble the nationwide gender proportions, with female slightly outnumbering male (53\%, compared to 51\% nationwide). Despite this shortcoming, we find a substantial correlation between several types of models using both the actual food names or the the caloric density of the foods mentioned in tweets and state-wide obesity and diabetes rates. This suggests that, in aggregate, social media does provide useful insights into national dietary health, even though the underlying data might not be representative.


It is an advantage of big data analysis that high-precision methods, such as hand-crafted keyword filters, can be effectively applied to glimpse a phenomena of interest. 
Yet, such approaches are not robust under temporal changes in the vocabulary, and may suffer from low recall. 
For example, the \texttt{\#fatproblems} hash tags are a convention for people to admit (often humorously) to feeling overweight. 
However, there may be many more subtle (and more serious) ways one can detect self-image expressions. 
Being able to identify with high accuracy users that are overweight would allow a more fine-grained validation of our techniques, 
in addition to the state-level validation that we currently focus on.

Despite the significant relationship between food mentions and their caloric value in the tweets and health problems, 
one needs to be careful not to assume that the user consumed every food about which they tweeted. In fact, it is difficult to extrapolate whether the tweet is about an actual dining experience, even if we detect a mention of food. A crowd-sourced effort to label a training set of tweets for detecting dining experiences has shown the task to be difficult, with user agreement at 78\%, and the resulting trained classifier producing noisy output. We leave determining the exact nature of the food mention in a tweet to future work.

Another concern is whether the tweets are about foods not normally eaten by their writers, that is, written on special occasions or on special topics. Thus, we checked the intervals at which the users in our dataset tweet, finding the top 1\% tweeting at a median of 18.8 food tweets per week, and the median overall to be at 1.2 food tweets per week. This suggests that our users at least do not tweet just about their birthday cake, but are actively engaged on a weekly basis. Although in our analysis we normalize our data per user, we still may have unusual very-active users who tweet differently than the rest. Thus, we check the overlap between the top mentioned food between the top 1\% and the bottom 50\%, and find 43 out of 50 foods overlap. Finally, even the top foods we find are every-day ones, like \emph{pizza}, \emph{coffee}, and \emph{chicken}. These findings encourage us that we capture at least in part the day-to-day dietary habits of Twitter users.

Our analysis describes correlations, not causations. However, we believe that insights gained from this type of analysis are required before deciding where to drill deeper through, ideally, controlled experiments. Targeting Twitter users with a particular behavior could also be a promising step for interventions such as public health awareness campaigns.

Finally, active promotion of dietary habits were studied by Yom-Tov et al.\ \cite{yom2012pro}, who track the dissemination of pro- and anti-anorexia photos on Flickr and the emerging social networks. They find the two groups to interact mostly within each respective community, but for pro-recovery group to tag their content with terms which would ensure their content is visible to pro-anorexia users. Such fine-grained analysis is beyond the scope of our current study, but the direct social influence in terms of verbal and non-verbal interactions is an enticing future direction of this research.

\section{Conclusions}
\label{sec:conclusion}

In this paper we describe a large-scale study of the Twittersphere, as it allows us to monitor US-wide nutritional behavior. We show that the foods mentioned in the daily tweets of users are predictive of the national obesity and diabetes statistics, with values of $r=.77$ and $r=.66$ across the 50 US states and the District of Columbia. We show how the calories tweeted are linked to user interest and demographic indicators, and that users sharing more friends are more likely to display a similar interest toward food. More needs to be done to develop sensitive and accurate tools for user characterization, with both textual and social network information available. As a documentation of users' interests, opinions, and behaviors, this study is another example of the potential Twitter has for public health research.

{
\bibliographystyle{acm-sigchi}
\footnotesize
\bibliography{twitterfood}
}
\balance
\end{document}